\documentclass[reprint, superscriptaddress, showkeys]{revtex4-1} 

\usepackage{graphicx}
\usepackage{amsmath}
\usepackage{amssymb}
\usepackage{hyperref}

\newcommand{\beq}{\begin{equation}}
\newcommand{\eeq}{\end{equation}}
\newcommand{\bea}{\begin{align}}
\newcommand{\eea}{\end{align}}

\newcommand{\ga}{\gamma}
\newcommand{\la}{\lambda}
\newcommand{\ze}{\zeta}

\newcommand{\ka}{\kappa}

\newcommand{\nn}{\nonumber}
\newcommand{\pa}{\partial}

\newcommand{\B}{\big}
\newcommand{\BB}{\bigg}

\begin{document}

\title{Blast wave kinematics: theory, experiments, and applications}
\author{Jorge S. D\'\i az}
\affiliation{Physics Department, Indiana University, Bloomington, IN 47405, U.S.A.}
\author{Sam E. Rigby}
\affiliation{Department of Civil and Structural Engineering, University of Sheffield, Mappin Street, Sheffield S1 3JD, UK}

\date{October 14, 2021}

\begin{abstract}
Measurements of the time of arrival of shock waves from explosions can serve as powerful markers of the evolution of the shock front for determining crucial parameters driving the blast.
Using standard theoretical tools and a simple ansatz for solving the hydrodynamics equations, a general expression for the Mach number of the shock front is derived.
Dimensionless coordinates are introduced allowing a straightforward visualization and direct comparison of blast waves produced by a variety of explosions, including chemical, nuclear, and laser-induced plasmas.
The results are validated by determining the yield of a wide range of explosions, using data from gram-size charges to thermonuclear tests.
\end{abstract}
\maketitle

\section{Introduction}
\label{Sec:intro}

Recent large-scale industrial accidents such as those in Tianjin (2015; 173 deaths) and Beirut (2020; 218 deaths) provide stark illustrations of the devastating potential of explosions. 
In addition to the tragic loss of human lives, the latter caused an estimated \$15B in property damage: complete destruction of buildings extended to a few hundred metres from the source of the explosion, and broken glass and debris was observed at distances up to 3~km from the explosion center, encompassing an area with more than 750,000 inhabitants \cite{Agapiou2020}. Clearly, in order for engineers to design structures for resilience against explosions, the properties of the blast wave must be known both relatively close to (where the structure should be designed to avoid/limit progressive and disproportionate collapse) and relatively far from the source (where the majority of injuries are caused by either lacerations from airborne glass fragments or by damage to hearing from failed glass panels \cite{Norville1999}).

Knowledge of the arrival time of a blast wave at various distances from the source enables a radius-time relationship to be developed, from which other key parameters such as peak pressure can be derived \cite{Dewey1964,Dewey1971}. Thus, the ability to determine this relationship {\it a priori}, from a known explosive yield, will provide vital information on the properties of the blast wave as it propagates. Further, a well-defined relationship that is valid for any distance permits the yield of an explosive to be determined through inverse analysis \cite{Gallet2021}.

This article presents a description of the propagation of a shock wave produced by an explosion in free air, an extension of the standard strong-shock solution to its later phase transitioning into an acoustic wave, and the applications of the results for estimating the yield of a wide variety of explosions as well as  the method is outlined for its future application.

\section{Theoretical description of the blast wave}
\label{Sec:blast-wave}

Let us model the shock wave produced by an explosion in free air as a sphere of time-dependent radius $R$.
A reflection factor can be used to extend the results in this section to explosions in the vicinity of surfaces and those produced by hemispherical charges.
The energy $E_0$ of the explosion is assumed to be released instantaneously and in a minuscule volume in air of undisturbed ambient conditions of atmospheric pressure $P_0$ and density $\rho_0$.
Conservation of energy and the equation of state of an ideal gas can be used to write the energy released by the explosion in terms of the kinetic and thermal energy of the gas contained within a radius $R$ in the form
\begin{align}
E_0 &= 4\pi\int_0^R \left(\frac12 \rho u^2+\frac{P-P_0}{\ga-1}\right)r^2 dr, \label{E0}
\end{align}
where $r$ represents a radial coordinate measuring the distance from the center of the explosion to the shock front $R$.
The factor $\ga$ is the heat capacity ratio, assumed to be unaffected by the passing of the shock;
its value for air in normal conditions described as a diatomic gas is $\ga = 1.4$. 
The radial velocity $u$, pressure $P$, and density $\rho$ of the air behind the shock front satisfy well-known hydrodynamics equations, which must be solved to determine their radial dependence before performing the integration in (\ref{E0}).

The PDE system describing the motion, continuity, and equation of state of the fluid are respectively given by
\begin{align}
\frac{\pa u}{\pa t} + u\frac{\pa u}{\pa r} &= -\frac{1}{\rho}\frac{\pa P}{\pa r}, \label{eq-u}\\
\frac{\pa\rho}{\pa t} + u\frac{\pa \rho}{\pa r} + \rho\left(\frac{\pa u}{\pa r}+\frac{2u}{r}\right)&= 0, \label{eq-rho}\\
\left(\frac{\pa}{\pa t} + u\frac{\pa}{\pa r}\right)\left(P\rho^{-\gamma}\right) &= 0, \label{eq-P}
\end{align}
subject to the boundary conditions at the shock front given by the Rankine-Hugoniot relations.
In terms of the Mach number of shock front $M_S=a_0^{-1}(dR/dt)$, these relations are
\begin{align}
u(R)    &= \frac{2a_0M_S}{\ga+1}\B(1-M_S^{-2}\B), \label{BC-u}\\
\rho(R) &= \frac{(\ga+1)\rho_0}{\ga-1+2M_S^{-2}}, \label{BC-rho}\\
P(R)    &= \BB(\frac{2\ga M_S^2 - (\ga-1)}{\ga+1} \BB) P_0,\label{BC-P}
\end{align}
where $a_0=\sqrt{\ga P_0/\rho_0}$ is the speed of sound at ambient conditions.
Let us characterize the motion of the shock front by introducing the dimensionless variables
\beq
\eta = \frac{r}{R}, \quad
\la  = M_S^{-2}, \label{eta,la}
\eeq
where $\eta$ specifies the distance from the explosion center ($\eta=0$) to the shock front  ($\eta=1$); whereas $\la$ characterizes the speed of the shock front from high Mach number ($\la\to0$) to the ambient speed of sound ($\la=1$).
Let us now write the ratios of the three quantities of interest in terms of the new variables as
\beq
\frac{u}{a_0M_S}       = \phi(\eta, \la), \quad
\frac{\rho}{\rho_0} = \psi(\eta, \la), \quad
\frac{P}{P_0}     = \frac{f(\eta, \la)}{\lambda}. \label{phi-psi-f}
\eeq
Note that for the strong-shock regime ($M_S\gg1$) Taylor's definitions \cite{GITaylor-I} are recovered.
Using the functions (\ref{phi-psi-f}), the energy equation (\ref{E0}) can be rewritten as
\beq
z^{-3} + K_1 = \la^{-1} K(\la), \label{E0(K)}
\eeq
where we have introduced the dimensionless scaled distance $z=R/R_0$, which measures distance in units of the explosion characteristic length $R_0 = (E_0/P_0)^{1/3}$.
Notice that $z$ differs from the standard scaled distance $Z=R/W^{1/3}$ used in blast engineering;
the latter normalizes the distance by the cubic root of the mass $W$ of the explosive charge, whereas $z$ removes cumbersome units and, more importantly, eliminates sometimes problematic TNT equivalence of different explosive materials.
The function $K(\la)$ is defined as
\beq
K(\la) = 4\pi \int_0^1 \bigg(\frac{\ga}{2}\psi(\eta,\la)\phi^2(\eta,\la) + \frac{f(\eta,\la)}{\ga-1} \bigg)\eta^2 d\eta. \label{K(la)}
\eeq
In the limit $R\to\infty$ the blast wave decays to an acoustic wave ($\la\to1$), hence the constant $K_1\equiv K(1)$ in (\ref{E0(K)}) corresponds to the boundary value of $K$ in the far field.
The decay of the blast wave can be parametrized by the auxiliary function
\beq
\zeta(\la) = \frac{R}{3\la} \frac{d\la}{dR}, \label{zeta}
\eeq
that describes how the speed of the shock front decreases as it moves away from the explosion center.
From the energy equation (\ref{E0(K)}), it follows that this auxiliary function and $K(\la)$ are related by the ordinary differential equation
\beq
K - \la K_1 = \ze(\la)\BB(K-\la \frac{dK}{d\la}\BB).\label{K(zeta)}
\eeq
This relation implies that the auxiliary function must satisfy the boundary conditions $\ze(0) = 1,  \ze(1) = 0$.
The simplest description of the blast decay that allows for an analytical description of the Mach number of the shock front and satisfies the boundary conditions is the linear decay $\ze(\la) = 1 - \la$.
Numerical analysis and experimental observations suggest that the decay is nonlinear, with the Mach number decaying more rapidly at early times.
In this work we intend to provide an approximate description of the phenomena; therefore, the linear choice will suffice.
Since the boundary conditions are satisfied, our approximate description will match the exact solutions in the early and late regimes, whereas some small deviation can appear in the mid-range where the strong shock transitions to the acoustic wave.
In Sec. \ref{Sec:experiments} we will see that the linear ansatz provides an accurate description of the blast wave for all ranges.

The linear form of the auxiliary function $\ze(\la)$ leads to a simple solution of (\ref{K(zeta)}) given by
\beq
K(\la) = (1-\la)K_0 + \la K_1,
\eeq
where the integration constant has been chosen so that $K_0$ denotes $K(\la)$ evaluated at $\la=0$.
This solution allows inverting the energy equation (\ref{E0(K)}) to write the the Mach number in terms of the scaled distance as
\beq
M_S(z) = \frac{dz}{d\tau} = \BB(1 + \frac{1}{K_0z^3}\BB)^{1/2}, \label{M(z)}
\eeq
where we have introduced the dimensionless scaled time $\tau = a_0t/R_0$.
Another reason for using these dimensionless variables $(\tau,z)$ is that they allow direct comparison of a wide range of experiments independent of the yield of the explosion under consideration. This enables us to visualize the results from gram-sized charges to megaton yields from thermonuclear explosions in the same plot, as is done in this article.
Notice that $\la$ in (\ref{eta,la}), the auxiliary function (\ref{zeta}), and its linear form can also be used to write a nonlinear differential equation for $M_S(z)$.
This equation is of the Bernoulli type so that it can be analytically solved; its solution is again given by (\ref{M(z)}), which confirms the mathematical self-consistency of the system.

\begin{figure}
\includegraphics[width=0.48\textwidth]{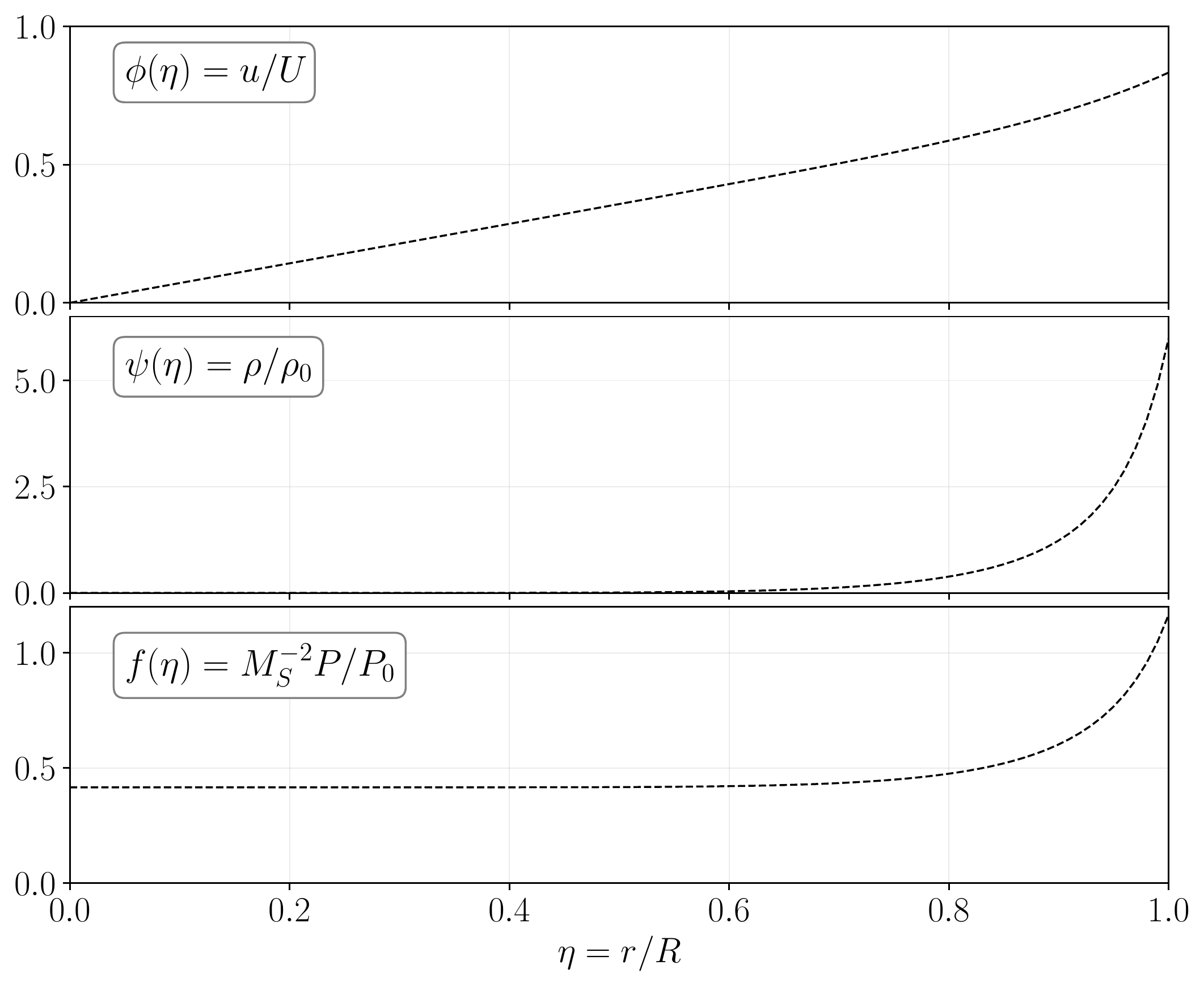}
\caption{Solutions of the hydrodynamics equations as functions of the dimensionless radial coordinate $\eta$.} 
\label{Fig:hydrodynamic-functions}
\end{figure}

\begin{figure}
\includegraphics[width=0.48\textwidth]{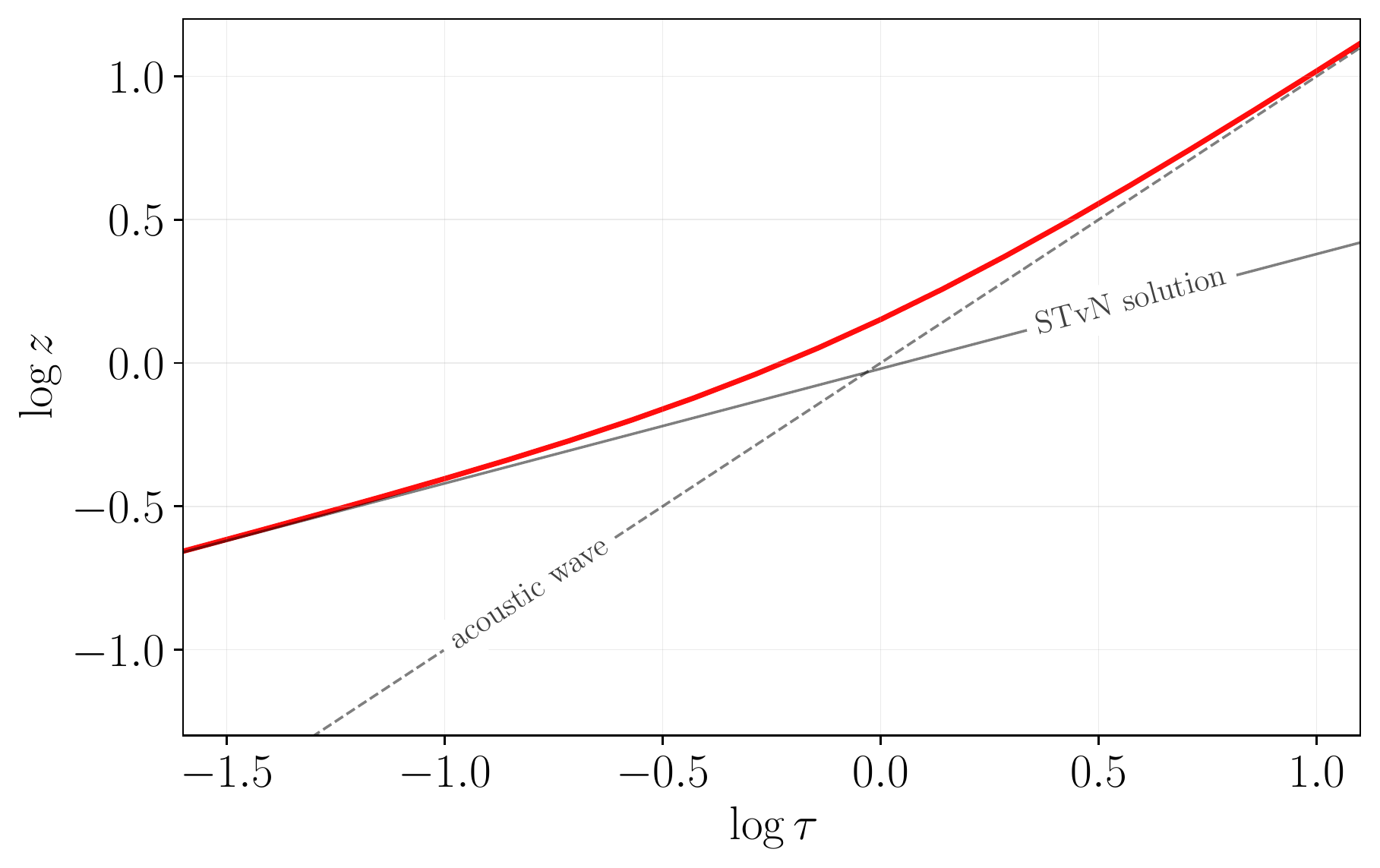}
\caption{Blast-wave solutions: the solution of (\ref{M(z)}) smoothly transitions from the STvN solution (see Sec. \ref{Sec:STvN}) to the acoustic regime characterizing the decay of the blast wave to an acoustic wave.} 
\label{Fig:blast-solutions}
\end{figure}

The solution for the Mach number (\ref{M(z)}) shows that only the numerical value of the function $K(\la)$ (\ref{K(la)}) at $\la=0$ is needed for fully describing the propagation of the shock front.
This observation in turn implies that the solutions of the hydrodynamics functions $\phi(\eta,\la), \psi(\eta,\la)$, and $f(\eta,\la)$ are necessary only at $\la=0$, which significantly simplifies the ODE system (\ref{eq-u}--\ref{eq-P}).
Using the definitions (\ref{phi-psi-f}), the solution to the system (\ref{eq-u}--\ref{eq-P}) with boundary conditions given by the Rankine-Hugoniot relations (\ref{BC-u}--\ref{BC-P}) at $\la=0$ is
\begin{align}
\phi(\eta,0) &= \frac{\eta}{\ga} + \BB(\frac{\ga-1}{\ga^2+\ga}\BB)\eta^{\ka_1}, \label{phi(eta)}\\
\psi(\eta,0) &= \BB(\frac{\ga+1}{\ga-1}\BB) \frac{\eta^{\ka_2}}{\ga^{\ka_3}}\B(\ga+1-\eta^{\ka_1-1}\B)^{\ka_3}, \label{psi(eta)}\\
f(\eta,0) &= \BB(\frac{2\ga^{1-\ka_4}}{\ga+1}\BB)\B(\ga+1-\eta^{\ka_1-1}\B)^{\ka_4},
\end{align}
where the exponents $\ka_i$, $i=1,\ldots,4$ are only functions of the heat capacity ratio $\ga$:
\begin{align}
\ka_1 &= \frac{7\ga-1}{\ga^2-1}, \quad
\ka_2 = \frac{3}{\ga-1}, \quad
\ka_3 = \frac{2\ga+10}{\ga-7}, \nn\\
\ka_4 &= \frac{2\ga^2+7\ga-3}{\ga-7}.
\end{align}
The three functions in terms of the dimensionless radial coordinate are shown in 
Fig. \ref{Fig:hydrodynamic-functions}.
%
We can now use these solutions in the definition of $K(\la)$ to determine $K_0$ in the form
\begin{align}
K_0 &= 4\pi \int_0^1 \bigg(\frac{\ga}{2}\psi(\eta,0)\phi^2(\eta,0) + \frac{f(\eta,0)}{\ga-1} \bigg)\eta^2 d\eta \nn\\
&= 7.86,
\end{align}
where the heat capacity ratio for air has been used since we have assumed the explosion to take place in free air.
Once this value is determined, the Mach-number equation (\ref{M(z)}) can be used to describe the growth of the spherical shock front as a function of the distance from the explosion center.
The general solution of (\ref{M(z)}) is shown in Fig. \ref{Fig:blast-solutions} together with the strong-shock solution discussed in Sec.~\ref{Sec:STvN} and the acoustic wave that the general solution must asymptotically approach.

Given the analytical form of the Mach number (\ref{M(z)}), the Rankine-Hugoniot relations can be used to write a simple expression for the peak hydrostatic overpressure behind the shock front as
\begin{align}
\Delta P &= \frac{7P_0}{6}(M_S^2-1) = \frac{7P_0}{6K_0z^3} 
= \frac{7E_m}{6K_0} Z^{-3}, \label{dP}
\end{align}
where the last form is relevant for chemical explosions. 
The energy of the explosion has been related to the mass of a charge by $E_0=E_mW$, where $E_m$ is the specific energy per unit mass that characterizes the chemical energy converted into kinetic and thermal energy after the explosion.
For example, considering a TNT explosion ($E_m\approx4.3$ MJ/kg) the oversimplified expression (\ref{dP}) leads to an overpressure barely distinguishable from the Brode formula for spherical blasts \cite{Brode1955}.

\section{Sedov-Taylor-von Neumann Blast Wave}
\label{Sec:STvN}

The famous Sedov-Taylor-von Neumann solution \cite{Sedov, GITaylor-I, JvNeumann} assumes a strong shock ($P\gg P_0$), which corresponds to setting $\la=0$ and neglecting the thermal energy of the air before the explosion.
This is equivalent to solving the blast-wave equation (\ref{M(z)}) for the early stages of the explosion when $z^3\ll K_0^{-1}$, simplifying the Mach number equation to the reduced form
\beq
M_S(z) = \frac{dz}{d\tau} \approx K_0^{-1/2} z^{-3/2}, \label{dz/dtau-STvN}
\eeq
whose solution is
\beq
z(\tau) = \BB(\frac{25}{4K_0}\BB)^{1/5} \tau^{2/5}, \label{STvN-soln-ztau}
\eeq
shown in Fig. \ref{Fig:blast-solutions}  as a straight line of slope 2/5 in the log-log plane.
In standard coordinates, we recover the more familiar form
\beq
\frac{dR}{dt} = \BB(\frac{\ga E_0}{K_0\rho_0}\BB)^{1/2} \,R^{-3/2}, \label{eq-STvM}
\eeq
whose solution is the well-known STvN blast wave
\beq
R = \BB(\frac{25\ga}{4K_0}\BB)^{1/5}\BB(\frac{E_0t^2}{\rho_0}\BB)^{1/5}.
\eeq
The constant factor for air is
\beq
S(\ga) = \BB(\frac{25\ga}{4K_0}\BB)^{1/5} = 1.022,
\eeq
which is moderately closer to the exact value $S(1.4)=1.033$ than the approximate result $S(1.4) = 1.014$ found by Chernyi \cite{Chernyi}.
It should be emphasized that this description of a blast wave is only valid in the early stages of expansion and where the explosion can be assumed to originate as point-source energy release, such as a nuclear explosion or in the mid-range for a chemical explosion.
In a later stage, a blast wave will decay and the strong-shock approximation will no longer be valid (and in the early stages of a chemical explosion the energy release will not be from a point-source).
For a full description of the blast wave, and more crucially, including the transition from a string shock to an acoustic wave we must solve the equation for the general Mach number (\ref{M(z)}).


As shown in (\ref{dz/dtau-STvN}), the STvN solution is obtained when neglecting the thermal energy of the undisturbed air before the explosion via the strong shock condition ($P\gg P_0$).
Similarly, by comparing the general differential equation (\ref{M(z)}) describing the blast wave and the STvN limit (\ref{dz/dtau-STvN}), we can write an upper value for the validity of the STvN solution from the general expression (\ref{M(z)}) in the form
\beq
z_\text{upp} \lesssim K_0^{-1/3} = 0.50.
\eeq
For scaled distances higher than $z_\text{upp}$ deviations from the STvN solution are expected due to the decay of the shock wave.
This behavior is independent from the type of explosion; chemical or nuclear.

\begin{table}
\caption{Range of validity of the STvN solution for some explosives.
Values of $E_m$ from \cite{LLNL-handbook, AN-Em}.}
\label{table:Z_minmax}
\begin{tabular}{lccc}
\hline\noalign{\smallskip}
	&	TNT	&	PE4/C4	&	AN	\\
\noalign{\smallskip}\hline\noalign{\smallskip}
$E_m$ (MJ/kg)	&	4.294	&	5.621	&	1.447	\\
$z_\text{low}$	&	0.21	&	0.19	&	0.30	\\
$z_\text{upp}$	&	0.50	&	0.50	&	0.50	\\
$Z_\text{low}$ (m/kg$^{1/3}$)	&	0.73	&	0.73	&	0.73	\\
$Z_\text{upp}$ (m/kg$^{1/3}$)	&	1.77	&	1.94	&	1.23	\\
\noalign{\smallskip}\hline
\end{tabular}
\end{table}

In the other direction, there is also a lower value $z_\text{low}$ for the range of validity of the STvN solution for chemical explosions.
The solution neglects the mass of the explosive charge $W$ compared to the mass of the surrounding air over which energy has to be transferred.
For this reason, there is a minimum distance from the center of the explosion where the mass of the charge can no longer be neglected.
Imposing the condition $m_\text{air} \gtrsim 2W$, we find 
\beq
\BB(\frac{3P_0}{2\pi\rho_0 E_m}\BB)^{1/3} \lesssim z_\text{low},
\eeq
where $E_m$ is the specific energy per unit mass introduced in the previous section. 
In standard dimensions, the range of validity of the STvN solution can be written in the form
\beq
\BB(\frac{3W}{2\pi\rho_0}\BB)^{1/3} < R < \BB(\frac{E_m W}{K_0P_0}\BB)^{1/3}.
\eeq
Using scaled distance $Z=R/W^{1/3}$, the range of validity of the STvN solution in air becomes
\beq
0.73 \text{ m/kg}^{1/3} < Z < \big(1.3\, E_m\big)^{1/3} \text{ m/kg}^{1/3} ,
\eeq
where the specific energy per unit mass $E_m$ must be in MJ/kg.
Explicit values for TNT, PE4, and ammonium nitrate are presented in Table \ref{table:Z_minmax}.

\section{Experiments}
\label{Sec:experiments}

As mentioned in the previous section, a very nice property of the dimensionless scaled coordinates ($\tau,z$) is that we can visualize explosions from multiple different yields in a single plot.
In this section we consider measurements of the arrival time of the shock front at different distances for a variety of explosions and show how these measurements agree with the results from the previous sections.
For nuclear explosions, the units kt and Mt refer to $10^3$ and $10^6$ tons of TNTe, respectively.

\begin{figure}
\centering %
\includegraphics[width=0.48\textwidth]{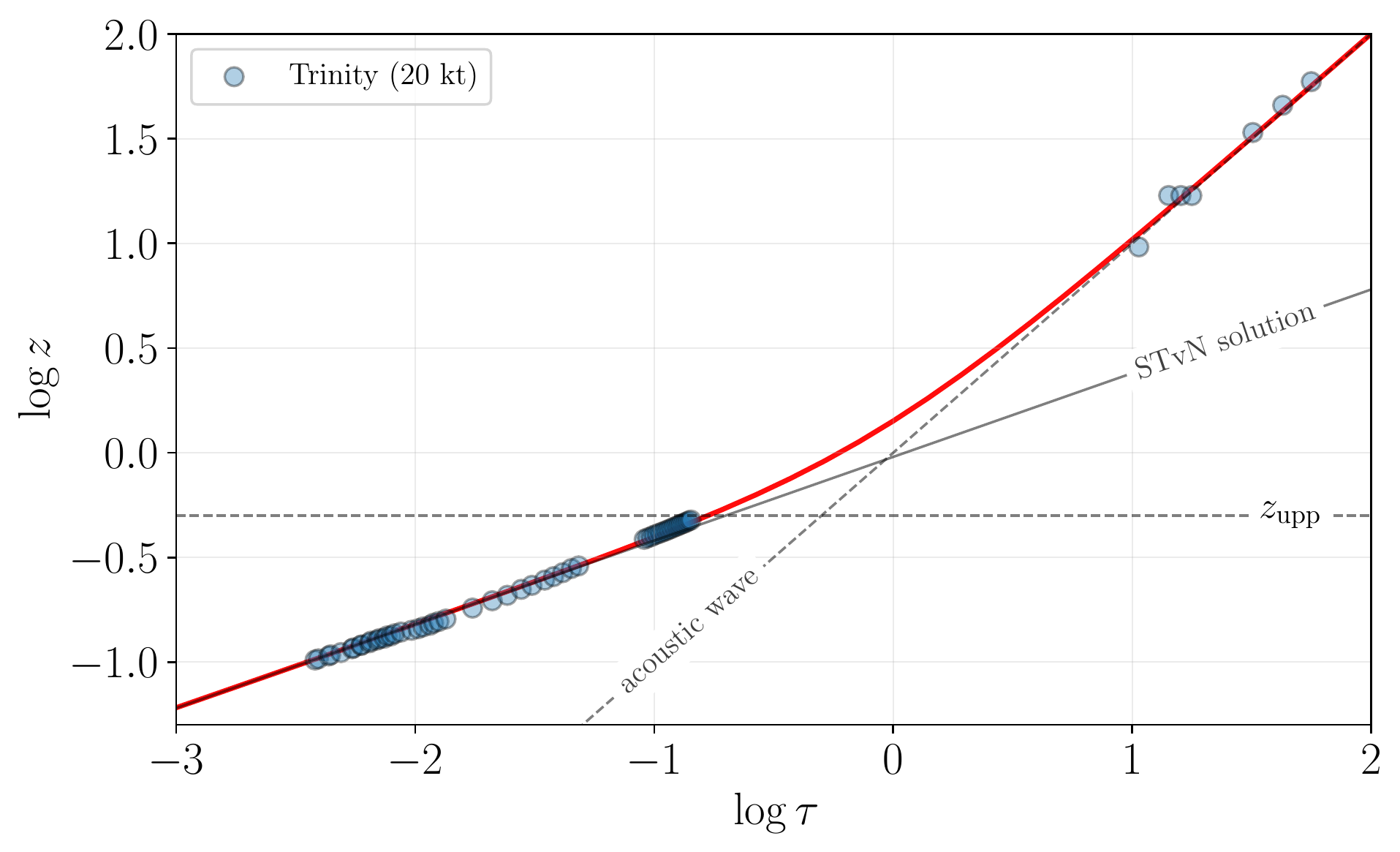}
\caption{Blast-wave data of the Trinity test. 
As described by Taylor \cite{GITaylor-II}, the fireball data follows the STvN solution. 
The measurements reported by witnesses of the test from different locations follow the curve in the acoustic regime.} 
\label{Fig:Trinity}
\end{figure}
\subsection{Gram-sized explosive charges}

\begin{figure*}
\centering %
\includegraphics[width=\textwidth]{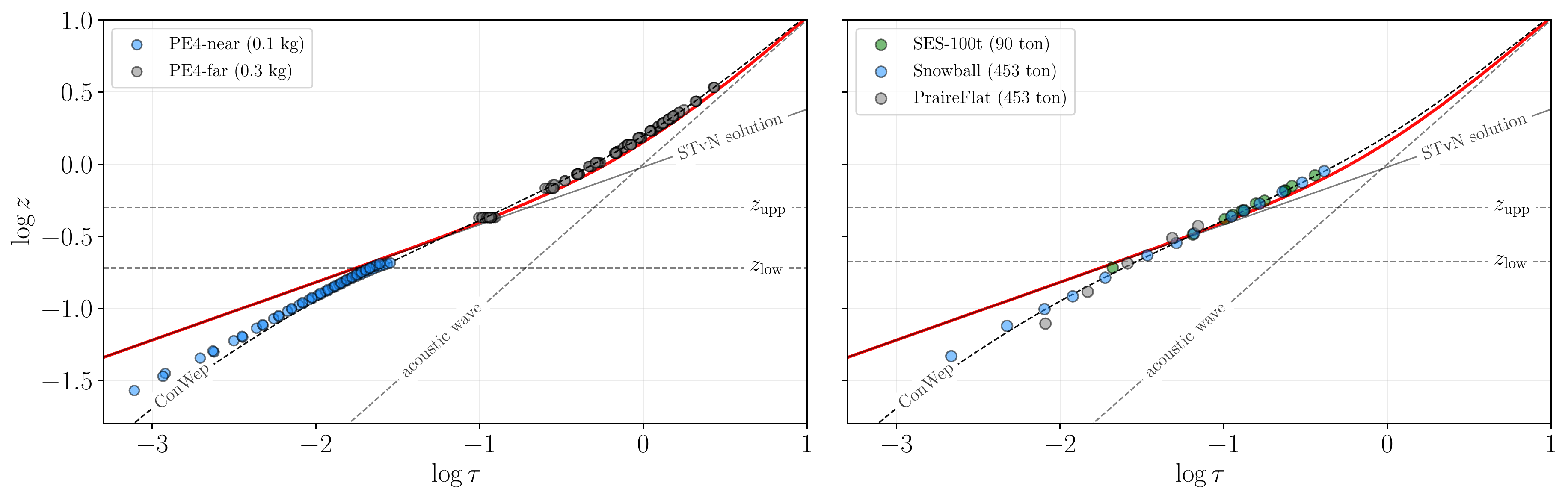}
\includegraphics[width=\textwidth]{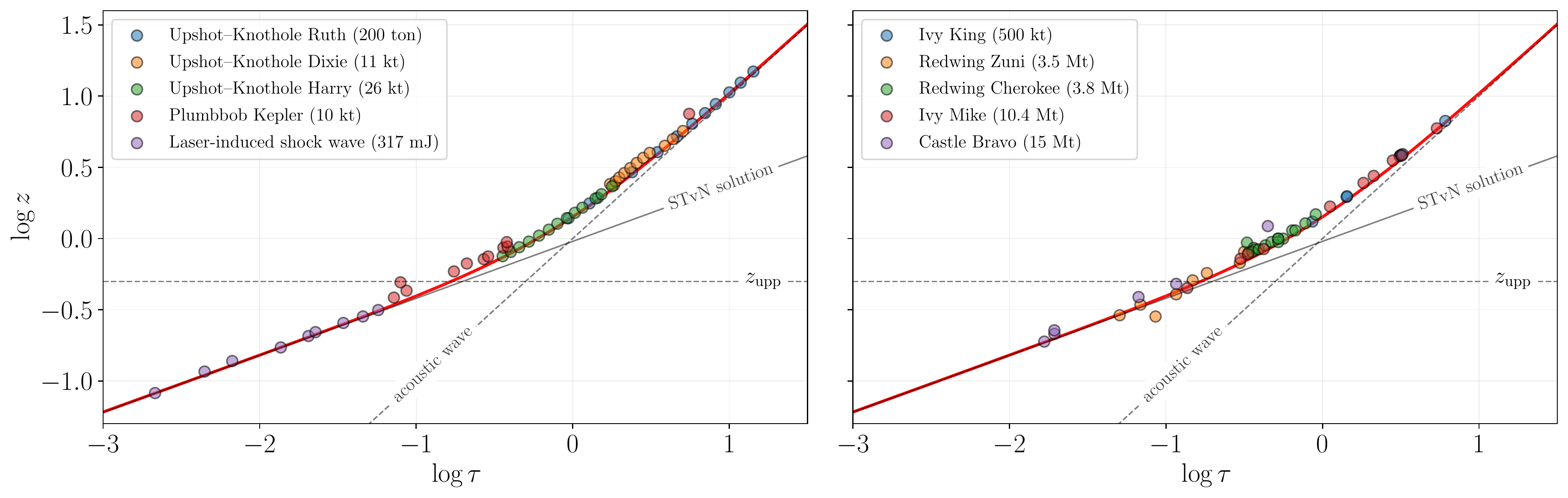}
\caption{Top: Blast-wave data from a collection of gram-sized charges (left) and large chemical explosions (right). The upper and lower limits for the validity of the STvN solutions are indicated and the curve of the ConWep data for 1 kg of TNT is also shown.
Bottom: Blast-wave data from a collection of early nuclear tests and from laser-induced shock waves (left); and data from some historical thermonuclear tests (right).} 
\label{Fig:chemical_nuclear_explosions}
\end{figure*}

The explosion of gram-sized charges offer the possibility of studying the very early stages of a blast as well as the influence of different charge geometries.
High-speed cameras allow for recording of the early shock wave and sensitive devices can measure the overpressure without being destroyed by the blast. In recent years, researchers at the University of Sheffield (UoS) Blast and Impact Laboratory have conducted approximately 80 far-field arena tests using hemispheres of PE4 explosive \cite{Rigby2014far,Rigby2014neg,Rigby2015AOI,Tyas2011,Tyas2019}, and a smaller number of near-field tests using spheres of PE4 \cite{Rigby2015near,Rigby2020HSV}.
The results are shown in the top-left panel of Fig. \ref{Fig:chemical_nuclear_explosions}. 
For comparison, the figure also includes the curve of the ConWep data for 1 kg of TNT \cite{ConWep}. 

\subsection{Large chemical explosions}

Many tests of significant amounts of explosives have been carried out using TNT and ANFO to mimic the effects of kiloton-range nuclear explosions \cite{Snowball, Prairie-Flat}.
Accidental explosions, such as the Beirut blast \cite{Pilger, Rigby, Diaz}, also allow for studies in this range.
The data for a selection of explosions in this range is shown in the top-right panel of Fig. \ref{Fig:chemical_nuclear_explosions}. 

\subsection{Early nuclear explosions}

From the first nuclear test (Trinity), nuclear explosions with yields in the dozens of kilotons were abundant during the late 1940s through to the 1950s.
Many unclassified technical reports of these tests include information of the pressure measurements at different distances from ground zero \cite{Upshot-Knothole, Plumbbob}.
In particular, Trinity is the only test for which early data is available and this is in fact what G.I. Taylor used in his second paper \cite{GITaylor-II}; however, the far-field data is missing.
General Leslie Groves requested many firsthand accounts describing the reactions of people who witnessed the Trinity test \cite{TrinityWitnesses}.
The reports by the scientists include information of their location and arrival time of the blast wave, which we have used to map the evolution of the Trinity blast in the far-field region
shown in Fig. \ref{Fig:Trinity}. 
For all later nuclear tests only mid- to far-field data is available, whereas early-time measurements at millisecond scales remain unpublished.
A team of scientists, historians, and filmmakers at Los Alamos and Livermore National Laboratories are currently working on the restoration and digitization of old nuclear-tests films and it is expected that fireball data will be published in the near future \cite{Carr}.

\subsection{Laser-induced shock waves}

Shock waves can be generated by the fast deposition of energy in different materials by laser pulses.
A second laser can be used for diagnostics of the produced plasma and some of its properties can be inferred by studying the time evolution of the shock as well as the plasma plume in a variety of geometries.
These laser-induced shocks are usually characterized by the STvN solution \cite{LISP}.
Data of a spherical shock produced by a joule-range laser is included in the bottom-left panel of Fig. \ref{Fig:chemical_nuclear_explosions}. 

\subsection{Thermonuclear explosions}

During the Cold War the development of advanced nuclear weapons pushed the yield from kilotons to megaton thermonuclear tests \cite{Ivy, Redwing, Castle}.
The formidable amount of energy released by these explosions allow for reliable measurements only very far from ground zero; however, the high yields lead to short scaled distances and times into the mid-field region.
Results from a selection of thermonuclear test are shown in Fig. \ref{Fig:chemical_nuclear_explosions}.

\section{Applications}

\begin{figure}
\centering %
\includegraphics[width=0.48\textwidth]{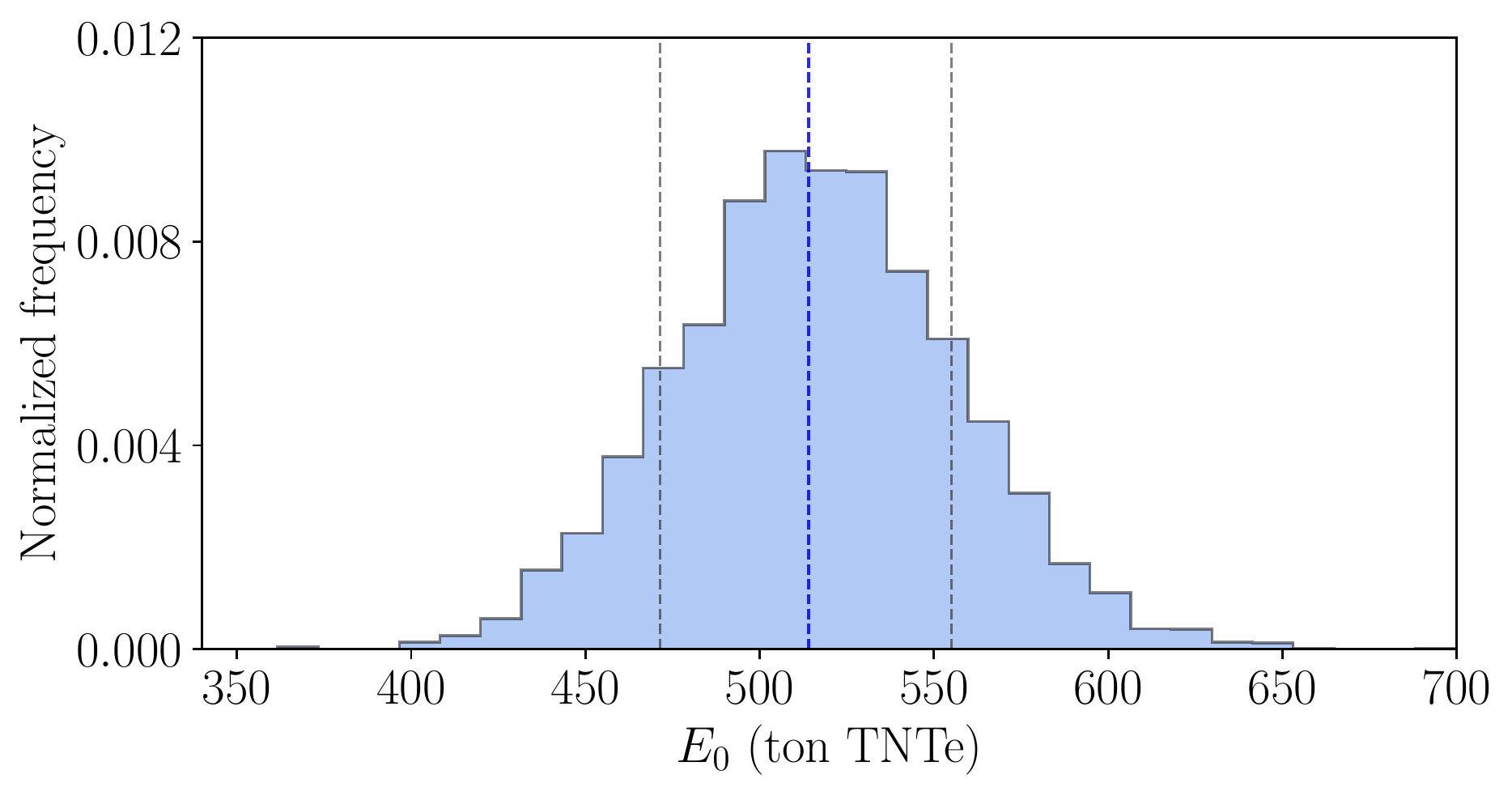}
\caption{Posterior probability distribution of the model parameter $E_0$. 
The value $E_0 = 514^{+41}_{-43}$ ton TNTe represents the median of the distribution and the uncertainties are based on the 16th and 84th percentiles of the sample, shown in the plot.} 
\label{Fig:Beirut-E0}
\end{figure}

One useful application of the results of the previous sections is the determination of the yield, $E_0$, of an explosion from a set of $(t,R)$ pairs.
It is tempting to simply fit the solution of Eq. (\ref{M(z)}) to data; 
nonetheless, 
there are a few considerations to keep in mind to avoid falling into conceptual traps:

\begin{enumerate}
\item One aspect to take into account is the behavior of the curve at different ranges.
As shown in Fig. \ref{Fig:blast-solutions},
the solution coincides with the STvN line in the short range, meaning that for very early times and short distances the solution might fail to properly describe a chemical explosion; this is not an issue for nuclear explosions, as mentioned in Sec.~\ref{Sec:STvN}.

\item Additionally, the solution in the long range asymptotically approaches the acoustic wave ($M_S\to1$) independent of the energy $E_0$. This feature translates into a  highly degenerate solution, making the use of long-range-only data unreliable for determining $E_0$. 
This degeneracy is broken in the short range, and for this reason short-range data is crucial for a reliable determination of $E_0$.

\item Fitting the solution of Eq. (\ref{M(z)}) to data in the $(t, R)$ space makes the analysis highly sensitive to the values of the long-range data,
where large uncertainties can render the analysis useless.
Furthermore, the breaking of degeneracy described in the previous point is negligible on a linear scale.
Instead, the fit ought to be carried out in the $(\log\tau, \log z)$ space, where the log-log scale eliminates the problems from the linear scale.

\item A consequence of using the $(\log\tau, \log z)$ space for the fit is that the individual uncertainties (measured in the $(t, R)$ space) become large in the short range and small in the long range.
\end{enumerate}

\begin{figure}
\centering %
\includegraphics[width=0.48\textwidth]{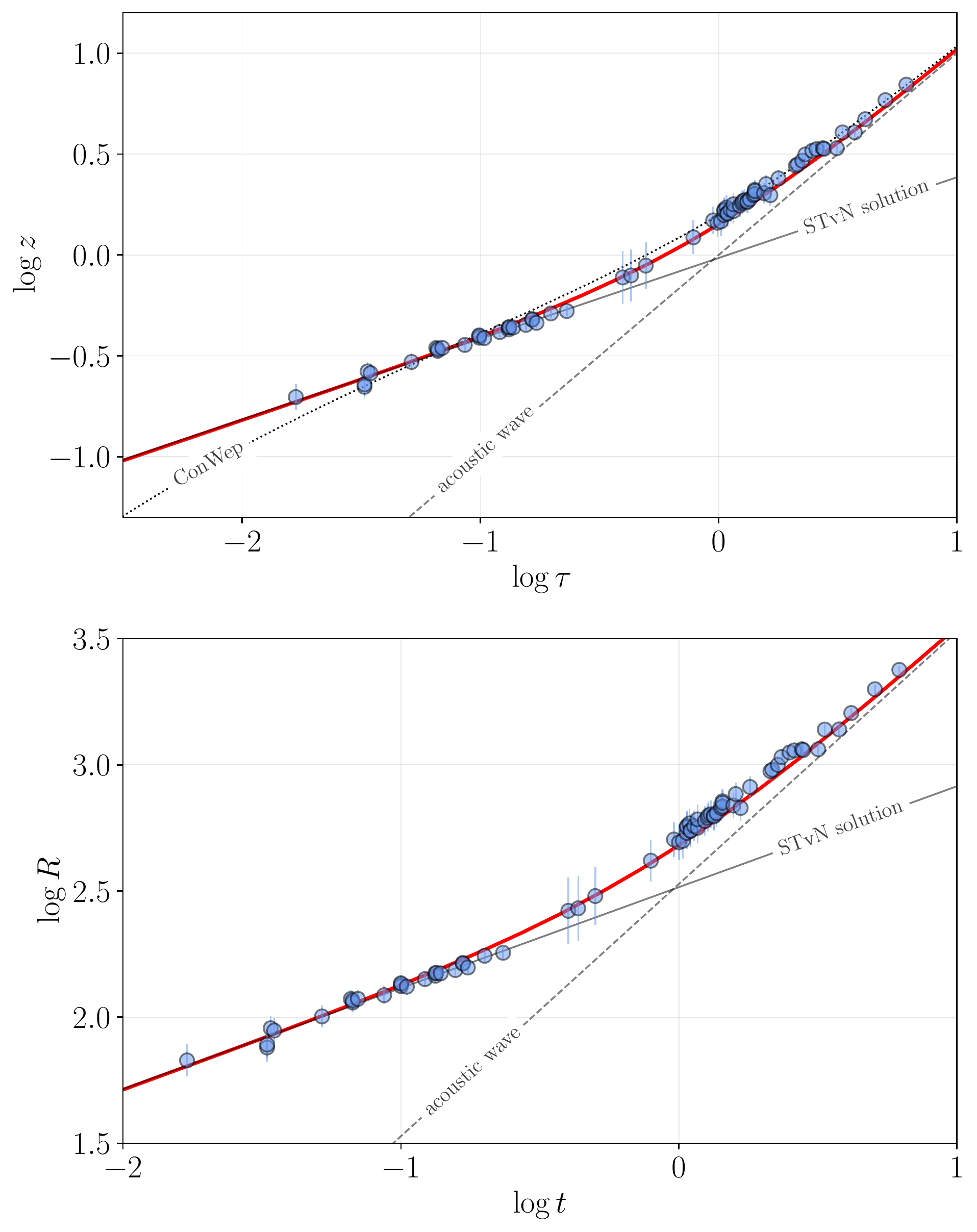}
\caption{Data from the Beirut explosion; the value $E_0 = 514$ ton TNTe was used for scaling the data (top) in the dimensionless coordinates ($\tau,z$) and the blast-wave curve (bottom) in standard coordinates ($t,R$).} 
\label{Fig:Beirut-plots}
\end{figure}

\begin{figure*}
\centering %
\includegraphics[width=0.98\textwidth]{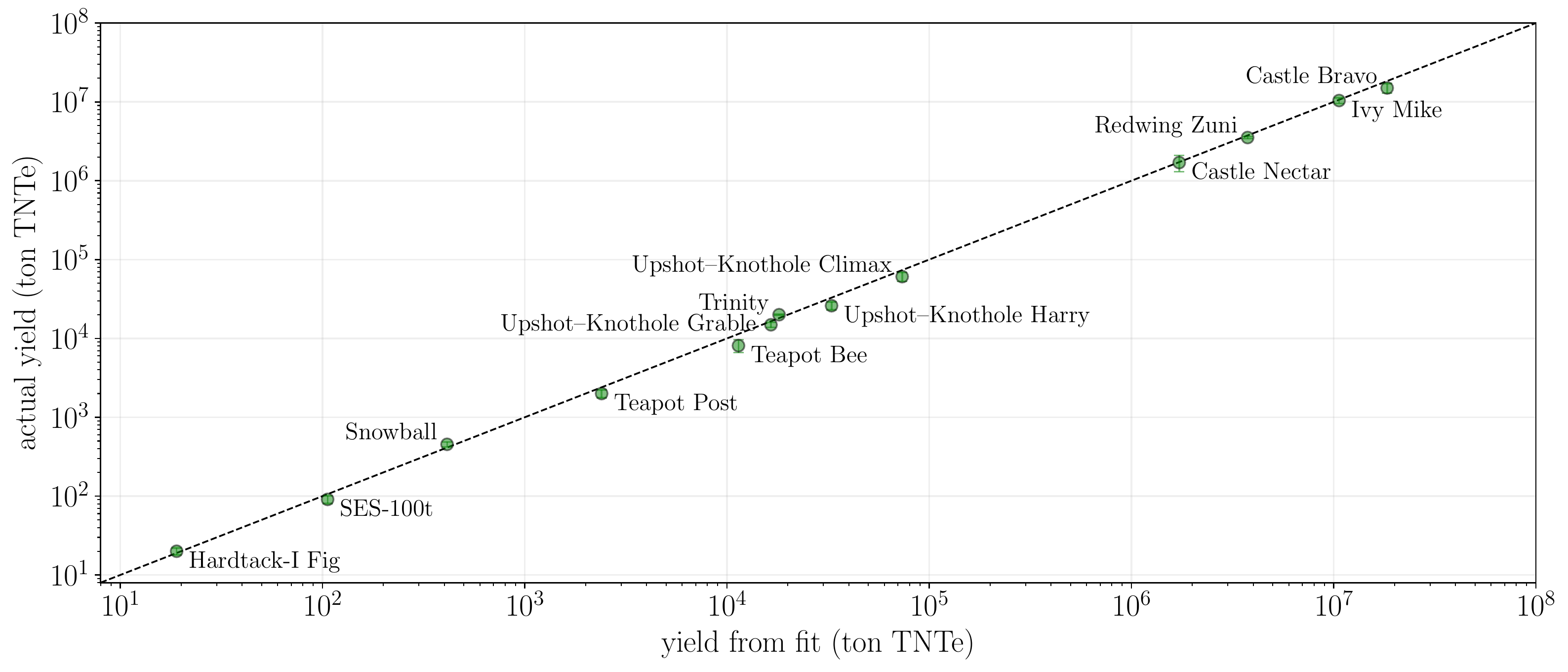}
\caption{Comparison between fitted and actual yield $E_0$ for thirteen explosions over a wide range of energies shown in units of tons TNTe.} 
\label{Fig:fit-vs-actual-yields}
\end{figure*}

As an illustrative example, let us consider the data set of $(t,R)$ pairs from the Beirut explosion \cite{Rigby, Diaz} and use the results from Sec.~\ref{Sec:blast-wave} to estimate the yield that caused this blast.
We can relate the physical quantities $t$ and $R$ to the dimensionless variables $\tau$ and $z$ using the unknown parameter $E_0$ and then minimize a loss function with respect to the numerical solution of (\ref{M(z)}).
A robust method is obtained by using \texttt{emcee}, a Python implementation of the affine-invariant ensemble sampler for Markov Chain Monte Carlo (MCMC) \cite{emcee, emcee2}. 
Using the combined data sets from Refs. \cite{Rigby} and \cite{Diaz}, the resulting posterior probability distribution of the model parameter $E_0$ is shown in Figure \ref{Fig:Beirut-E0}.
The value $E_0 = 514^{+41}_{-43}$ ton TNTe represents the median of the distribution and the uncertainties are based on the 16th and 84th percentiles of the sample.
This value accounts for the fact that the Beirut explosion took place at ground level rather than in free air (assumed in previous sections).
The correction is obtained by dividing $E_0$ by the reflection factor 1.8 (for soil), to account for the enhancement of the shock wave due to the ground-reflected hemisphere and the energy loss due to cratering and ground shock \cite{factor18}. However, due to the built-up nature of the Port of Beirut and its surroundings, a factor of 1.8 is deemed appropriate.
For explosions near sea level, during nuclear tests on the Pacific Proving Grounds it was found that the reflection factor is closer to 1.6 due to extra energy dissipation in the form of large water displacements \cite{OperationCastle}.

Figure \ref{Fig:Beirut-plots} shows the data and the corresponding scaling using the value $E_0 = 514$ ton TNTe. 
Note that the first plot shows the $(\log\tau, \log z)$ space so the curves are scale independent, whereas the data is scaled.
On the contrary, the second plot shows the $(\log t, \log R)$ space, in which the curves rather than the data are scaled.

The value of $E_0$ found above is in excellent agreement with other yield determinations of the Beirut explosion, including those performed independently by each of the present authors \cite{Rigby, Diaz} and others.
As a more general validation of the method, we have applied it for estimating the yield of a  selection of high-explosives and nuclear tests over a wide range of energies, from a few tons of TNTe to the high yields of thermonuclear tests \cite{Snowball, Upshot-Knothole, Plumbbob, Ivy, Redwing, Castle, SES100t, HardtackI, Teapot}.
The results of the fit of $E_0$ for thirteen historical explosions are shown in Figure \ref{Fig:fit-vs-actual-yields}, where the fits are compared to the actual yield in tons of TNTe.
As indicated earlier, the accuracy of the parameter fit relies on the availability of data in the early stages. 
Similarly, the precision of the parameter fit depends on the noise level of the data set.
These features are noticeable in the figure for the noisiest data sets corresponding to the tests Bee (Operation Teapot) and Harry (Operation Upshot–Knothole).

The excellent agreement between the fitted and actual yields over several orders of magnitude confirms that (\ref{M(z)}) provides an acceptable description of the shock front, despite the unrefined approximation of a linear decay of the Mach number.
We remark in passing that an evident deviation from the exact description of the Mach-number appears as the decay into the acoustic regime according to (\ref{M(z)}) does not include the logarithmic dependency found both theoretically \cite{Bethe} and semi-empirically \cite{Dewey2016}.

\section{Summary and Conclusions}

This article illustrates the results of a general characterization of a blast wave in free air, extendable to other configurations by using a reflection factor.
A linear ansatz for the decay of the Mach number of the shock front as it expands allows for analytical solutions of the hydrodynamics equations that lead to a concise expression for the Mach number of shock front in terms of the distance from the explosion center.
Despite the unsophisticated approximation for the deceleration of the shock front, the subsequently obtained expressions show an excellent agreement with experimental data.

A simple formula for the Mach number was derived in the form of an ordinary differential equation, whose solution describes the position versus time development of the shock front.
Here is where time of arrival measurements can be used for estimating the energy released $E_0$, a crucial parameter that determines the shock evolution and the loading developed on obstacles with which it interacts.
The general solution found contains the well-known strong-shock solution as a limit in the early stage of the shock development, beyond this regime the solution describes the transition to an acoustic wave in the far-field.
Experimental data from gram-sized charges was used verify the validity of the results and later archival data from large-scale explosions was also employed using dimensionless coordinates for time and distance so that explosions from grams of PE4 to thermonuclear blasts can be visualized in a single diagram.
The solution found serves as a generalization of other descriptions of the decaying blast wave, in this case, valid from the early (strong) stage to the asymptotically acoustic behavior at the far field.

A discussion about the validity of the strong-shock solution was presented that can serve a valuable resource for blast engineers. 
The yield of over a dozen explosions was estimated as way to validate the results found in this work and the main aspects of a fit to time-of-arrival data are discussed.
Our results show that one of the key features when fitting the yield to time-of-arrival data is that this must be performed in a log-log space; otherwise, slight errors in far-field data will dramatically affect the estimate of $E_0$ and can possibly render the analysis useless.
This property is due to the highly degenerate nature of the blast-wave solution in far field, where all solutions asymptotically approach to an acoustic wave independent of the yield $E_0$.
Additionally, when applied to laser-induced shocks, the method outlined in this work becomes a direct diagnostic of the laser energy deposited in the material.

\begin{acknowledgements}
J.S.D. was supported in part by the Indiana University Center for Spacetime Symmetries.
He also acknowledges the delightful company of Dr. H. Fry and Dr. A. Rutherford with their {\it Curious Cases} during most of this work, and 
thanks J.C. Valenzuela for bringing laser-induced shock waves to our attention.

\end{acknowledgements}

\section*{Conflict of interest}
The authors declare that they have no conflict of interest.

\end{document}